# Gas electron multipliers with metal electrodes and their applications in detectors

B.M.Ovchinnikov*, V.V.Parusov

Institute of Nuclear Research of Russian Academy of Sciences, Moscow.

*Corresponding author; e-mail: ovchin@inr.ru

**Abstract**

Gas electron multipliers (GEMs) with wire (WGEMs) or metal electrodes (MGEMs), which don't use any plastic insulators between the electrodes are created. The absence of plastic insulation between the electrodes of WGEMs excludes unwanted leakage currents and spark breakdowns between the electrodes. Accidental spark events in such MGEM don't lead to their failure as far as positive ions quickly move away from the breakdown region by a strong electric field in the gap region. In this work, we present a review of such GEMs.

The chambers containing MGEMs (WGEMs) with pin-anodes are proposed as detectors for searching of interactions between Dark Matter particles and hydrogen ($H_2$). In this paper, we present a review of such chambers. For investigation of the gas mixtures (Ar+10% $H_2$, Ne+10%$H_2$, $H_2$ + 3ppmTMAE), the chamber containing WGEM with pin-anodes detection system was constructed. In this paper we present the results of an experimental study of these gaseous mixtures exited by an alpha source.

Finally, we discuss principles of operation of GEMs with pin-anodes as well as plans for constructing of large scale (150 mm x 150 mm) MGEM detectors.

## 1. Introduction

More than forty years ago G. Charpak and F. Sauli have introduced their Multi-Step Chambers to overcome limitations of gain in Parallel-Plate and Multi-Wire Proportional Chambers (MWPC) [1]. These MWPCs have revolutionized detection systems in high energy physics.

Currently there are different types of detectors for fast detection and localization of charged particles exist. One of them is a Gas Electron Multiplier (GEM). A standard Gas Electron Multiplier [2,3,4] consists of a thin composite sheet (plate) with two metal layers separated by a thin insulator and pierced by a regular matrix of open channels. These plates contain through holes on all their area, separation distances and diameters of which are approximately equal to the plate thickness (Fig.1). Inside these holes, which are filled with corresponding gases, in presence of strong electric fields, a multiplication of electrons takes place. GEMs provide the best spatial resolution and higher rate than the wire chambers (MWPC). More coarse macro-patterned detectors are thick-GEMs (THGEM) [5, 6, 7] or patterned resistive thick GEM devices (RETGEM) [8].
However, the most essential disadvantage of GEMs consists in their low reliability and stability. The matter is that in a process of dispersion of the GEM's cathode electrodes by positive ions of proportional avalanches in GEM with metal or high-resistive electrodes (RETGEM), a sedimentation of the sprayed carrying-out material on the walls of holes with subsequent leaks and breakdowns between electrodes takes place. It leads to subsequent decrease of the potential difference between GEM's electrodes and corresponding reduction of the multiplication factor in an avalanche.

Micro-pattern gaseous detectors (MPGD), due to their tiny electrode structure and small avalanche gaps, are very fragile and can be easily damaged by sparks appearing at high operational gains (typically at gains of $10^4$ or slightly more) [7].

Therefore, we were concentrated on development of more robust designs of GEM detectors with wire (WGEM) or metal electrodes (MGEM). The idea of WGEM without plastic insulators was first mentioned in our work [9]. In our subsequent works [10, 11] a MGEM with metal electrodes of diameter of 22 mm was designed and tested. In our next works [15-19, 21, 25] it was suggested that the search WIMP of Dark Matter with help of detecting system with GEMs can be performed. In the paper [13] we have described a novel concept of MGEMs.

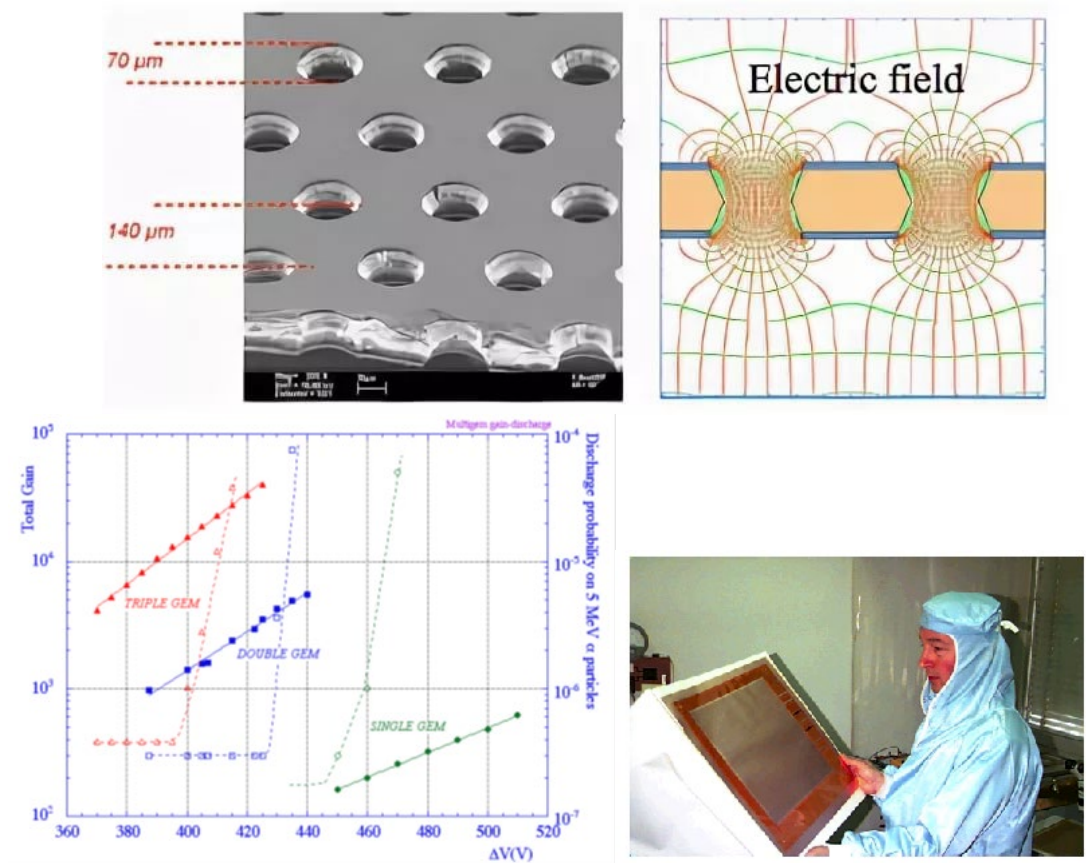

Fig.1 Gas Electrons Multiplier.

## 2. GEMs with wire electrodes (WGEM)

In our works [9, 10, 11] WGEMs with wire electrodes and no plastic insulators between them were created. The WGEMs [9] used macroscopic windows of size 1 mm by 1 mm, while WGEMs [10, 11] used windows of size 0,5 mm by 0,5 mm. The gap between the wire electrodes was equal to 1 mm. The design of wire GEMs and the results of their tests are shown in Fig.2 and Fig.3.

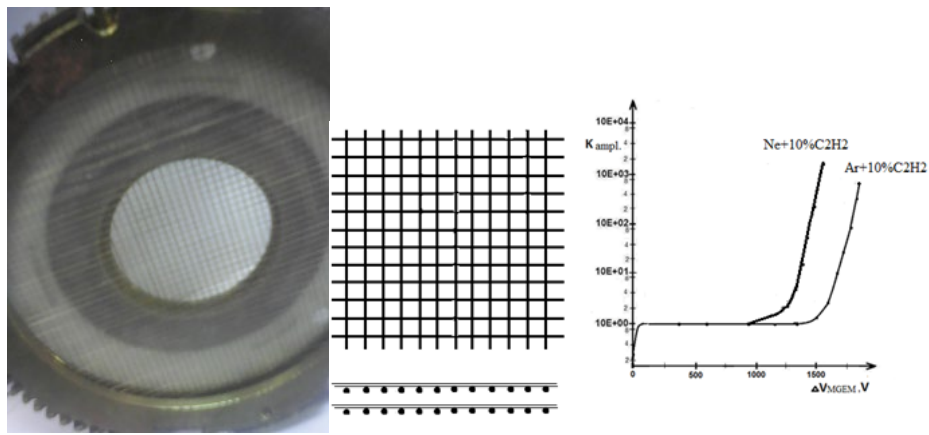

Fig.2. Design of WGEMs and results of their tests [9].

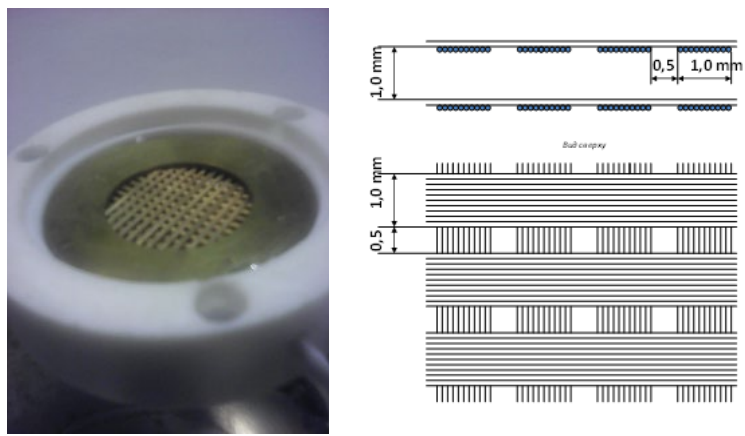

Fig.3. Design of WGEMs and results of their tests [10, 11].

## 3. GEM with metal electrodes (MGEM)

In the work [12] a MGEM with metal electrodes and sensitive area of 22 mm by 22 mm for the first time was demonstrated. The electrodes were made by drilling of 1 mm holes with a step between them of 1.5 mm in to 1 mm thick brass plates (see Fig.4). One disadvantage of that MGEM [12] is large duration of the process of drilling the holes in the electrodes, especially for a case of small diameters and small steps between holes. In addition to that, formation of agnails at the edges of holes in a process of drilling is possible. Another problem of drilling of holes with drilling machines consists in difficulty of production of large area MGEM electrodes with high accuracy of sizes of holes and their positions on various plates.

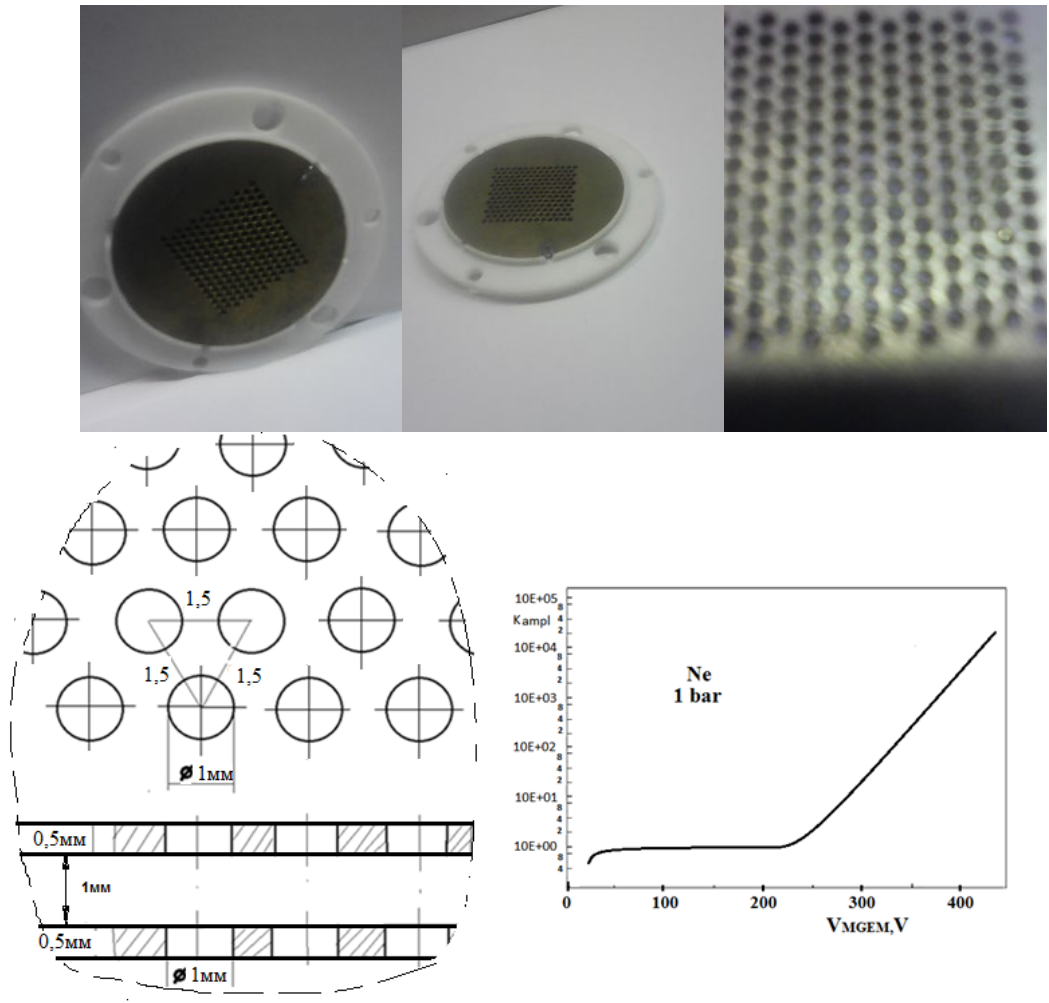

Fig.4. Design of metal GEM and results of their tests [12].

## 4. MGEM with the etching of holes

In the work [13] the gas electron multiplier with metal electrodes (MGEM), differing by its simplicity, high precision and technological effectiveness in production, as well as reliability and stability of its operation, is made and tested. To eliminate certain specific shortcomings (see Introduction), in this work metal electrodes were made by a method of drawing a mask on 0.3 mm thick brass plates with 0.3 mm diameter holes and 0.5 mm step between them (fig.5) with subsequent double-side etching of the holes.

As far as the initial brass plate was cut from 0.3 mm thick rolled foil, the produced electrodes had a curved shape. Therefore, at assembling, between the MGEM electrodes a fluoroplastic spacer was introduced to increase resistance on the pass of leakage charges between the electrodes. From the outer sides both electrodes were pressed to fluoroplastic plate by additional steel rings of 2 mm thickness.

By means of central 4 mm holes in electrodes and special insulating bolts, the GEM electrodes were mutually positioned in such a way that the relative displacements of holes didn't exceed 0.02 mm. The gap between electrodes was chosen to be equal to 1 mm. In this design a sensitive area GEM with holes had a diameter of D=75 mm. The GEM was tested in the chamber (Fig.6.) with various filling gases: Ar +10%$C_2H_4$ (1 and 0.4 atm), Ne +(($O_2$ +$N_2$ +$H_2O$)·$10^{-6}$)(1 and 0.4 atm), Ar +10% $CH_4$ (1 atm).

The results of tests at irradiation of the drift gap of the chamber by alpha-particles ($Pu^{239}$) are presented in Fig.5. It is visible, that Ne provides the maximum coefficient of multiplication at the smallest potential difference between electrodes before the breakdown happens.

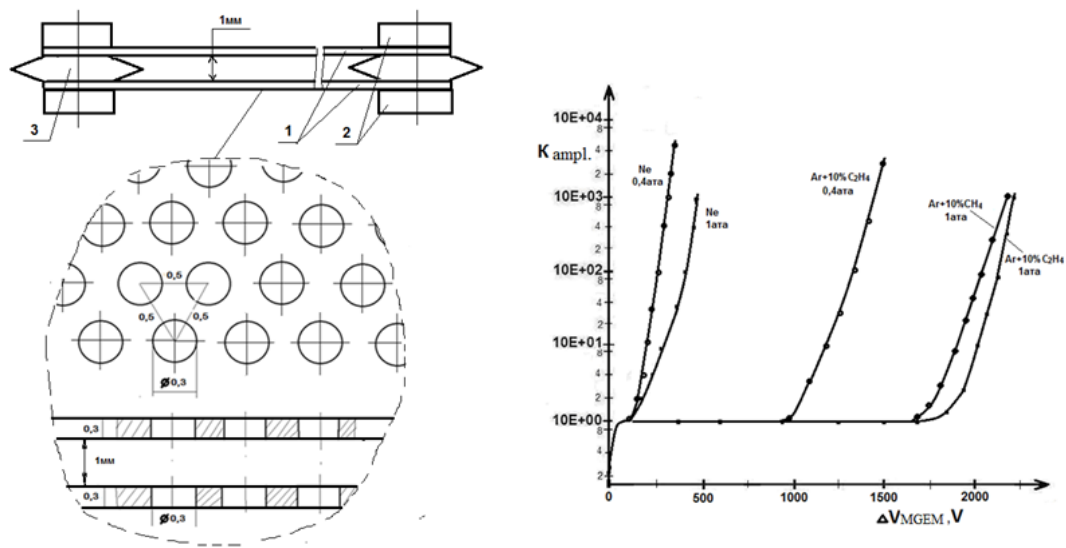

Fig.5. Left: design of MGEM: 1-plates of MGEM, 2-clamping rings, 3-layer from fluoroplastic. Right: amplification of MGEM filled with different gases.

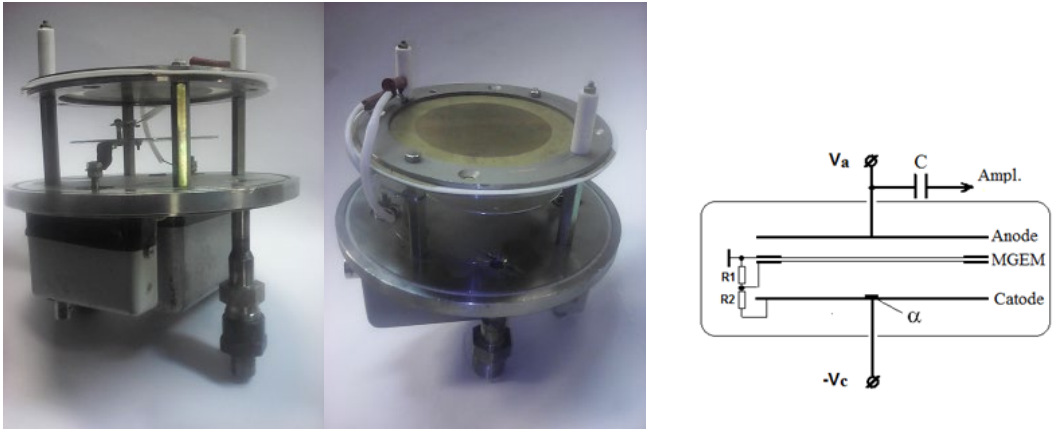

Fig.6. Chamber for test MGEM.

### 5. The detecting chambers with system GEM + pin-anodes for Direct Detection of WIMP

In work [15] the idea of focusing screen with holes + system pin-anodes already was shown (section 5.1). The wire gas electron multipliers in combination with pin-anodes are proposed for detection of events: in the gas phase of a double-phase argon chamber (section 5.2); in chamber for direct detection of WIMP with mass ≤ 0.5 GeV (section 5.3); and in chamber with gas mixture $H_2$ + 3ppm TMAE (section 5.4).

#### 5.1 A liquid-methane ionization chamber

A liquid-methane ionization chamber with a system of focusing screen + pin-anodes is proposed as a setup to search for spin-dependent interactions of DM particles with hydrogen [15]. The Anode of the chamber is placed in gaseous methane above liquid methane. The anode consists of a system of pins. The Focusing screen is placed between the Anode and liquid methane. The screen has a system of holes concentrically on the relevant pin-anode. The values of electrical potentials on the electrodes of the chamber are set in such a way that all electrical lines of force are focused on the pin-anodes. We provide only the idea of using a specially designed liquid-methane ionization chamber in an experiment aimed at searching for the (mostly) low mass DM based on their spin-dependent interactions with hydrogen.

#### 5.2 Double-Phase Argon Chamber

Multichannel WGEM + system pin-anodes are proposed for detection of events in the gas phase of a double-phase argon chamber [16, 17]. Hydrogen with a concentration of 10 % is added to argon to eliminate feedbacks via photons emitted by excited argon molecules in avalanche development processes during detection of events in the gaseous argon. A maximum electron multiplication coefficient of ~300 has been obtained for the multichannel wire gas electron multipliers with a 1 mm gap used to detect α-particles in the Ar + 10% $H_2$ mixture at a pressure of 1 bar. When a pin anode is used, the maximum electron multiplication factor for α-particles is ~$2.5 \times 10^5$. It has been experimentally shown that adding $H_2$ with a concentration of 100 ppm to liquid argon has no effect on the singlet component of the scintillation signal in the liquid argon and reduces the emission efficiency relative to the pure argon gas phase only slightly (by 20%).

To suppress the β, γ and n₀ backgrounds, the comparison of scintillation and ionization signals for every event is suggested [18, 19]. The addition in liquid Ar of photosensitive TMA, TMG or C2H4 [20] and suppression of triplet component of scintillation signals ensures the detection of scintillation signals with high efficiency and provides a complete suppression of the electron background.

### 5.3 The Chamber for Direct Detection of WIMP with Mass ≤ 0.5 GeV.

The chamber for direct detection of WIMP with mass ≤ 0.5 GeV was developed [21]. The chamber (see Fig.7) is filled with gas mixture Ne+10% Hydrogen +0,15ppm TMG. In this chamber for the events detection it was used a system GEM +pin-anodes, which provide the energy threshold about eV. The electron background is suppressed due to photosensitive addition of TMG. For a direct detection of WIMP it is proposed to use a liquid argon chamber with Hydrogen dissolved in liquid argon at a concentration 100ppm+0,015ppm TMG. Based on the work [22], where in a spherical proportional detector the energy threshold is about 100eV, while the amplification factor of the detecting system is about $10^4$, we estimate the threshold of our experiment to be about 100 eV·$10^4$/5·$10^7$ 1eV. The $H_2$-filling provides an efficient suppression of the electron background, because of the short track of recoil protons, compared to the one from background electrons [23].

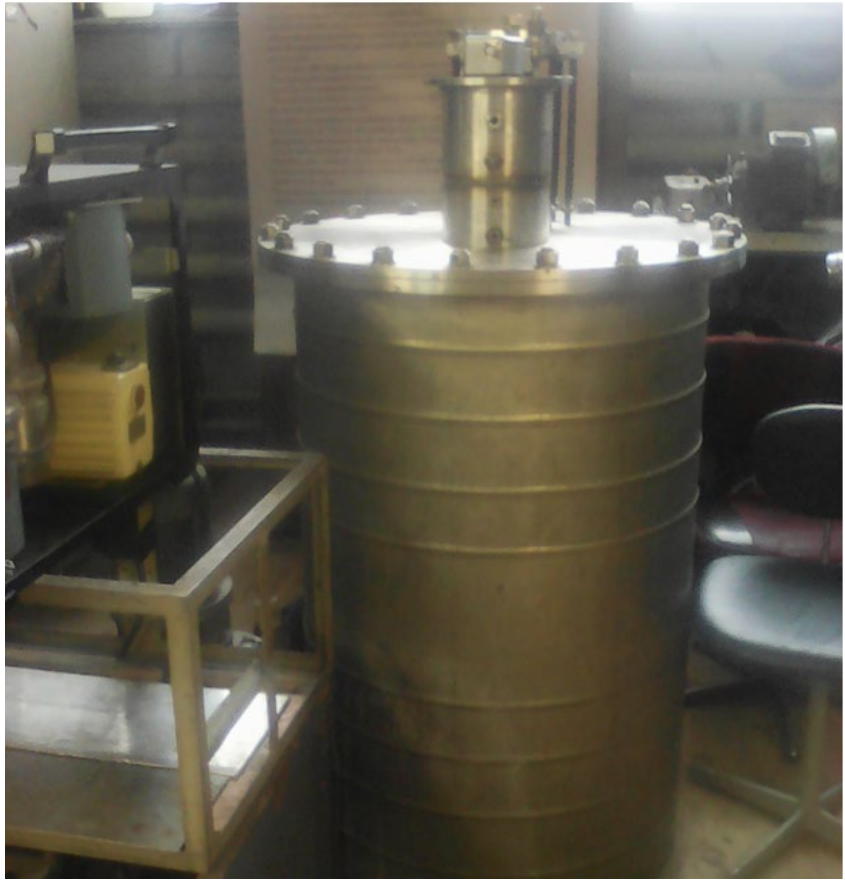

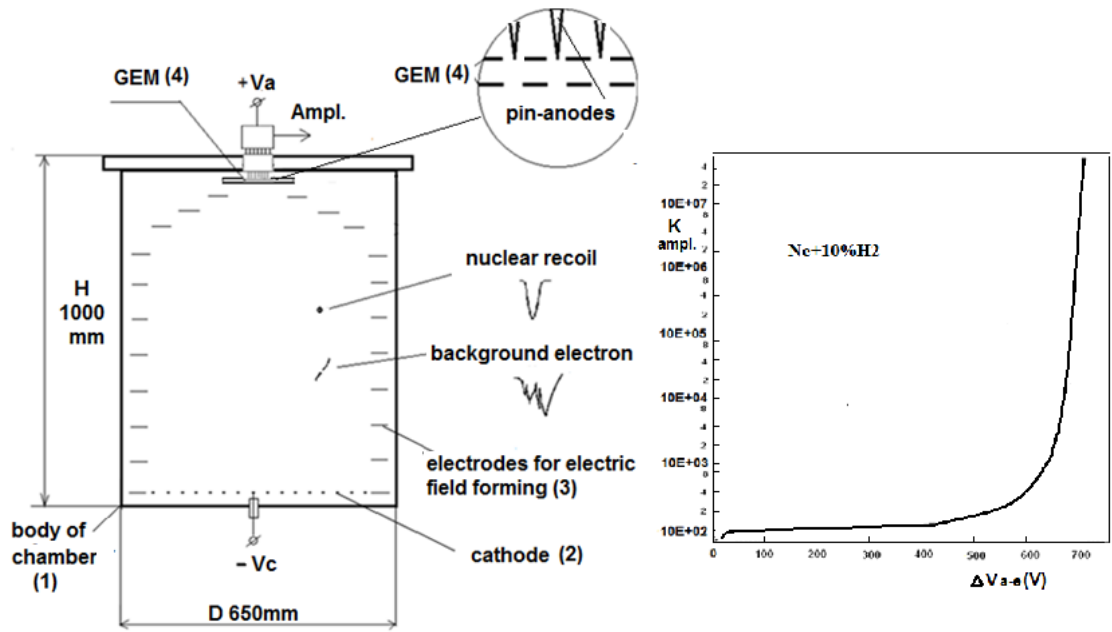

Fig.7. Chamber with a system GEM +pin-anodes and corresponding results of tests of the detection system (Fig.8). Coefficient of electron multiplication simultaneously in the WGEM (Kamp1 ~ $10^2$) and at the pin-anode (Kamp2 ~ $5·10^5$) as a function of the anode voltage. The chamber is filled with a mixture of Ne+10% at a pressure of 1 bar is exposed to α particles.

### 5.4 The chamber with gas mixture $H_2$ + 3ppm TMAE

The detectors with pure NaI, Xe or Ar [24, 26] make it possible to search WIMPs with large masses (up to dozens or hundreds GeV), as far as the energy of nuclear recoils in these detectors from low mass WIMPs is low. To account for yearly modulation effect in DAMA-LIBRA experiment [27] J.Va'vra has supposed [29] that this effect is explained by low mass WIMP scattered at protons in $H_2O$ molecules, which is contained in NaI crystals at about 1ppm level (see Table 1.)

In our work [25] the chamber for direct detection of WIMPs with mass < 10 Gev/$c^2$ and axions, emitted from the Sun, was developed. The chamber is filled with a gas mixture $H_2$ +3ppm TMAE (5, 10 bar), or $D_2$ + 3ppm TMAE. These gas fillings allow to suppress the electron background. For detection of events is used a system GEM + pin-anodes (Fig.8) with coefficient multiplication of about $10^5$ (see Fig.9 ) and the chamber of the previous experiment (section 5.3). Collisions of WIMPs with $H_2$ provide recoil protons with energies of several keV (see Table 1). An addition of TMAE with a low ionization potential (5,36 eV) provides detection of recoil protons.

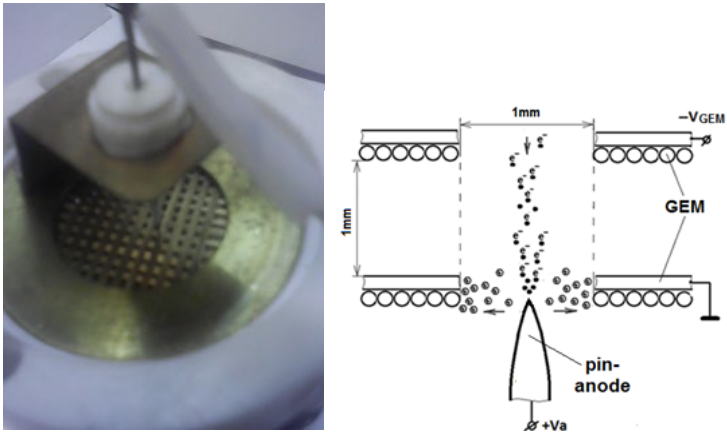

Fig.8. Detection system for testing of WGEM + pin-anode with diagram of the travel of positive ions from avalanches developed at the pin and electrons being collected at the pin.

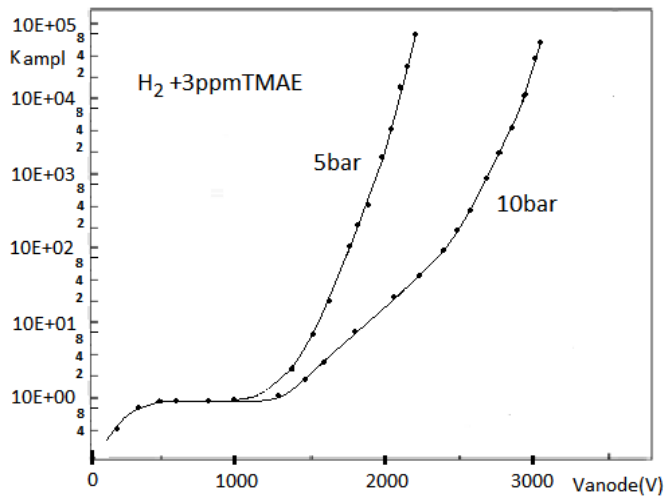

Fig.9. Measured amplification of a system GEM + pin-anode.

| WIMP [GeV/c²] | Nucl. | $E_{nr}$(keV) |
|---|---|---|
| 0,5 | H | 1,91 |
| 1,0 | H | 4,30 |
| 1,5 | H | 6,20 |
| 2,0 | H | 7,65 |
| 2,5 | H | 8,78 |
| 3,0 | H | 9,68 |
| 0,5 | Na | 0,19 |
| 1,0 | Na | 0,73 |
| 1,5 | Na | 1,57 |
| 4,0 | Na | 9,07 |

.

Table1. Maximum calculated nuclear recoil energy $E_{nr}$(keV) as a function of WIMP mass (GeV/c²) for two targets: hydrogen and sodium [29].

## 6. Operation principle of MWPC, GEM and system GEM+pin-anode

The operation principle of Multi-Wire Proportional Chambers (a), GEM (b) and system GEM + pin-anode (c) is illustrated in Fig.10. The factors which have allowed us to obtain high electron multiplication factors in the GEM + pin-anode system are as follows:

(1) High electric field strength in the system GEM + pin-anode makes it possible to obtain a big length of electron avalanche and high value of the electron multiplication factor ($10^6$-$10^7$);

(2) Positive ions from the avalanche at the pin are transferred by the electric field, mainly, to the walls of the hole in which the pin is located and, in smaller quantities, towards the ionization electrons being collected at the pin, which rules out the possibility of streamers being developed at the interface of the positive ion cloud and the electron avalanche [26];

(3) For GEM (see technology b) extraction efficiency decrease at low transfer fields values due to a worst electron extraction capability from the lower side of the GEM [26];

(4) Absence of a plastic insulation excludes the emergence of leakage current and spark breakdown between electrodes. Accidental spark events in such system don't lead to their failure as positive ions quickly move away from breakdown by a strong electric gap field.

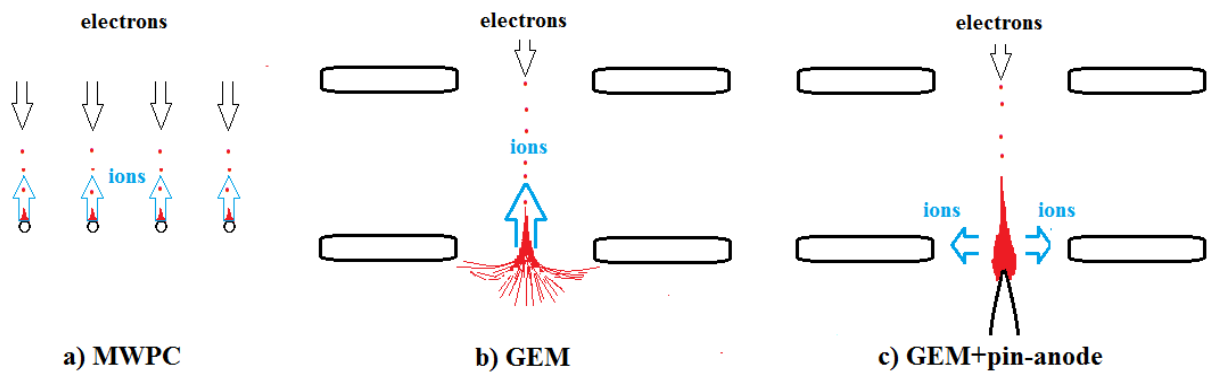

Fig.10. Operation principle of Multi-Wire Proportional Chambers (a), GEM (b) and GEM+pin-anode (c). Electron avalanches are shown for three technologies (a, b, c); red paths are electron trajectories, also the drift of ions is indicated (blue paths).

We have also plans for constructing of large scale (150mmx150mm) MGEM + pin-anodes detectors (see Fig.11).

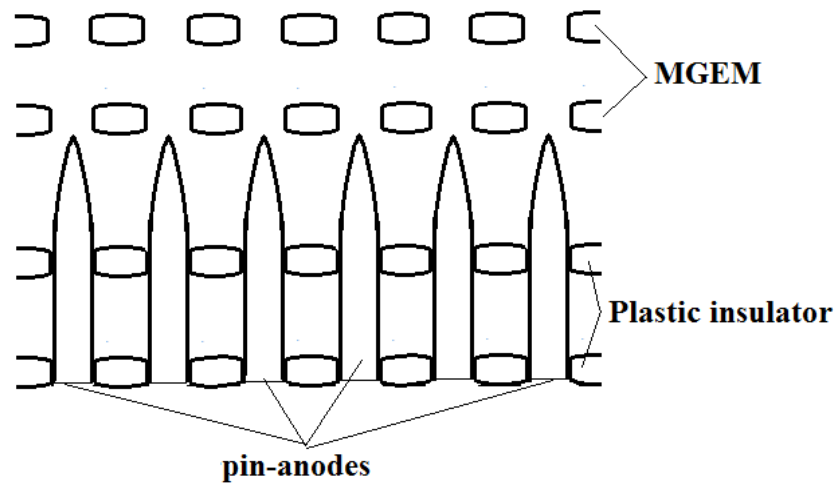

Fig.11. Schematic drawing of system GEM + pin-anodes in which plastic insulator contains of holes for pin-anodes. The metal electrodes (MGEM) made by a method of drawing a mask on brass plates.

## 7. Discussion

In works of CERN the multichannel gas electron multipliers consisting of a plastic plate 50-2000 microns thick with metal or high-resistive thin (some microns) deposits from two parties are presented [2-4]. These GEMs provide the best spatial resolution and higher rate that wire chambers. However, an essential disadvantage of these GEM consists in their low reliability and stability. The matter is that at cathode dispersion by positive ions from proportional avalanches in GEM with metal or high-resistive electrodes there is a sedimentation of the sprayed carrying-out material on walls of holes takes place, which leads to subsequent leaks and breakdowns between electrodes.

In our works GEMs with wire (WGEM) or metal electrodes (MGEM) and gas gap between metal electrodes without plastic insulators were created. Absence of a plastic insulation excludes the emergence of leakage current and spark breakdown between electrodes.

## 8. Summary & Outlook

In our works [9-14] GEMs with wire (WGEM) or metal electrodes (MGEM) and gas gap between metal electrodes without plastic were realized. An absence of a plastic insulation between electrodes of these GEMs excludes leakage currents and spark breakdowns between the electrodes.

In our works [15-17, 21, 25] it was suggested to search low mass WIMPs with help of chambers with GEMs and systems WGEM (MGEM) + pin-anodes. In that respect we would like to add the next important comments:

1. As far as WIMPs with large masses (> 10 GeV) experimentally were not found so far [24, 27, 28, 29], it is necessary to search the WIMP with small masses (≤ 10 GeV/$c^2$).

2. The data of the new DAMA/LIBRA–phase2 confirm a peculiar annual modulation of the single-hit scintillation events in the (1–6) keV energy range (WIMP mass < 10 GeV/c2) satisfying all of the multiple requirements of the Dark Matter [28]. J.Va'vra have supposed [29] that this effect is explained by low mass WIMP (~1 GeV/$c^2$) scattering on protons in $H_2O$ molecules (H+).

Finally, we would like to mention that MGEMs can have various applications in medicine. Such MGEMs can be used in different medical instruments for their use in X-ray surgery or Positron Emission Tomography (PET), where a high operation stability and reliability of the whole complex of instrument is required. Recently we have proposed a PET system, based on these GEMs [30].

## References.